\documentclass[sigconf,screen=false,review=false,anonymous=false,nonacm]{acmart}

\usepackage[yyyymmdd]{datetime}

\usepackage{tabularx}
\usepackage{dcolumn} %Aligning numbers by decimal points in table columns
\newcolumntype{d}[1]{D{.}{.}{#1}}

\usepackage{subcaption}

\usepackage{todonotes}
\let\xtodo\todo
\renewcommand{\todo}[1]{\xtodo[inline,color=green!50]{#1}}

\AtBeginDocument{%
  \providecommand\BibTeX{{%
    \normalfont B\kern-0.5em{\scshape i\kern-0.25em b}\kern-0.8em\TeX}}}

\setcopyright{CC}
\setcctype{by-sa}

\usepackage{todonotes}
\let\xtodo\todo
\renewcommand{\todo}[1]{\xtodo[inline,color=green!50]{#1}}

\begin{document}

\title{Leveraging Mobile Sensing Technology for Societal Change Towards more Sustainable Behavior}
%\subtitle{Opportunities and Ethical Considerations}

\author{Florian Bemmann}
\orcid{0000-0002-5759-4976}
\affiliation{%
  \institution{LMU Munich}
  % \streetaddress{P.O. Box 1212}
  \country{Germany}
}
\email{florian.bemmann@ifi.lmu.de}

\author{Carmen Mayer}
\affiliation{%
  \institution{LMU Munich}
  % \streetaddress{P.O. Box 1212}
  \country{Germany}
}
\email{C.Mayer@campus.lmu.de}

\author{Sven Mayer}
\orcid{0000-0001-5462-8782}
\affiliation{%
  \institution{LMU Munich}
  % \streetaddress{P.O. Box 1212}
  \country{Germany}
}
\email{info@sven-mayer.com}
% \affiliation{%

% \affiliation{%

% \affiliation{%

% \affiliation{%

% \affiliation{%

% \affiliation{%

% \affiliation{%

\renewcommand{\shortauthors}{Bemmann and Mayer, et al.}

\begin{abstract}
A pro-environmental attitude in the general population is essential to combat climate change. Society as a whole has the power to change economic processes through market demands and to exert pressure on policymakers - both are key social factors that currently undermine the goals of decarbonization.
% \cite{engels_anita_2023_11230}
% precise problem
Creating long-lasting, sustainable attitudes is challenging and behavior change technologies do hard to overcome their limitations. Environmental psychology proposes social factors to be relevant, a.o. creating a global identity feeling %\cite{reese2016common}
and widening one's view beyond the own bubble. %\todo{REF}. 
% What we do
From our experience in the field of mobile sensing and psychometric data inferences, we see strong potential in mobile sensing technologies to implement the aforementioned goals. We present concrete ideas in this paper, aiming to refine and extend them with the workshop and evaluate them afterward.
\end{abstract}






\maketitle

\section{Structure and Purpose of This Paper}

%\todo{6 pages max! (excluding references). Rather keep paper shorter (4-5) if possible}

% large scope problem
% A pro-environmental attitude in the general population is essential to combat climate change. Recent HCI efforts on behavior change technology, which we briefly give an overview on in the beginning of this paper, have faced their limitations in generating practical impact. However society as a whole has the power to change economic processes through market demands and to exert pressure on policy makers - which both are the key social factors that currently undermine the goals of decarbonization \cite{engels_anita_2023_11230}. Aft
% % precise problem
% Creating long-lasting, sustainable attitudes is challenging and behavior change technologies do hard overcoming their limitations. Environmental psychology proposes social factors to be relevant, a.o. creating a global identity feeling \cite{reese2016common} and widening one's view beyond the own bubble \todo{REF}. 
% % What we do
% From our experience in the field of mobile sensing and psychometric data inferences we see strong potential in mobile sensing technologies. We present concrete ideas in this paper, aiming to refine and extend them with the workshop and evaluate them afterwards.

% how this paper continues 
We will first briefly show how HCI has researched behavior change technology to support sustainable behavior and which limitations research is facing.
We then introduce research from behavioral- and environmental psychology, arguing that societal change and attitude forming are more promising than individual behavior change.
We show the powers of nowadays mobile sensing technology, data inference approaches, and social crowd sensing. We then bring the insights from environmental psychology and the presented HCI technologies together, to present novel application concepts based on these technologies that implement approaches that are promising to support societal change.
We discuss the proposed technology critically, as these technologies bring a high responsibility - mobile sensing data collection can raise severe privacy issues, and the application of psychometric targeting approaches is ethically critical and needs to be well discussed.
%\todo{incorporate? \citet{o2017roads}: The roads ahead: Narratives for shared socioeconomic pathways describing world futures in the 21st century}

\section{What SHCI recently did: Limitations of Behavior Change Technology}

Building on concepts of habit forming, self-optimization, and behavior change applications, HCI also investigated using such concepts to foster sustainable behavior, for example to push people more towards sustainable mobility \cite{froehlich2009ubigreen} or foster sustainable consumption through self-reflection \cite{bemmann2020self}.

%\todo{their limitations}
However persuasive sustainable interventions have limited real-world impact because the main objectives against acting sustainably are external circumstances that cannot be overcome by persuasive technology \cite{brynjarsdottir2012sustainably}. Furthermore, achieved behavior changes of studied projects are often not long-lasting in the wild \cite{hazas2012sustainability}. In their recent review, \citet{bremer2022have} summarize the efforts and limitations of past SHCI research, and call for going beyond individual behavior change and rather aim for societal change.

%---
Persuasive technologies in other domains are usually designed to directly improve an aspect of oneself (e.g. physical fitness, mental health) which can directly be tracked and an improvement be felt.
%\todo{[REF (optional): people feel more benefit of fitness apps because they directly inlfuence sth]}. Both success measures do not exist regarding sustainable behavior, as the aimed outcome (i.e. impact on climate change) cannot be quantified and seen directly.
Regarding sustainable behavior, classical behavior change-supporting technologies face limitations in real-world applicability, above all a lack of "good reason to use" e.g. extrinsic motivation, (see Technology Integration Model of \citet{shaw2018technology} for factors influencing continued use).

\section{Environmental Psychology + The Power of Societal Change}

Actual technology alone is not sufficient to combat climate change, societal change (that can be supported by technology) is at least as important \cite{engels_anita_2023_11230}. They report consumption patterns and corporate responses to be the two social factors that still undermine the goals of decarbonization. Hereby the latter is indirectly controlled by the first (i.e. companies adapt to market demands).
%\todo{literature on power-to-the-people, consumers have the power to infleunce companies, ...}
%\todo{from The "top 10 sozaile Faktoren" (siehe SZ / \citet{engels_anita_2023_11230}) And national politics (in my opinion the strongest) - also controlled by people decisions (at least in our country) -> read section 6.1.3}
Behavioral- and environmental psychology try to explain why people do not behave sustainably even though they have an attitude towards it (i.e. attitude-behavior gap), or what counteracts people developing an environmentally friendly attitude.

\paragraph{Attitude-Behavior Gap}

Regarding consumer behavior, the main barrier towards actual sustainable behavior are hard circumstances like price, perceived availability, and convenience \cite{aschemann2014elaborating}. A lack of such extrinsic motivational factors come together with rather weak intrinsic motivations: Moral short-sightedness \cite{ascher2006long} and %\todo{Zweifel ob man als einzelner etwas bewirkt [REF]} 
doubts whether one can make a difference as individuals throttle the intrinsic motivation of many people. Among the five obstacles towards far-sighted actions that \citet{ascher2006long} point out, especially selfishness and uncertainty play a role in our context. The effects of one's climate-negative actions are for western societies geographically far away (i.e. out of one's extended circle of selfishness) and the relationship is indirect, i.e. a concrete behavior does not directly lead to a concrete consequence.
%\todo{What specifically is \textit{moral} short-sightedness? -> Blink note}
% see philip hübl blinkist citation (google keep)}

Classical behavior change technologies (see e.g. \cite{fogg2009creating}) are thereby doing hard in making an actual change towards climate-friendly behaviors.

\paragraph{Pro-Environmental Attitudes}
In behavioral models, an attitude is a basis for behavior. Thus besides aiming for behavior change, the formation of a pro-environmental attitude among the population also is an important building block. \citet{reese2016common} argue that a common human identity, i.e. people regarding themselves as global citizens instead of part of some local group, could inform beliefs about environmental justice. \citet{huber2015gamification} propose instead to leverage the \textit{behavior-to-attitude link}. It is reported to be stronger than the vice versa link between attitude and behavior, although less studied yet. The behavior to attitude link can for example be observed when people are forced to life changes, e.g. when moving the location of home or workplace, in which associated higher flexibility towards pro-environmental change was observed \cite{weber2022office}.
%- global identity feeling \citet{reese2016common}. evlt auch: Global Citizens – Global Jet Setters? The Relation Between Global Identity, Sufficiency Orientation, Travelling, and a Socio-Ecological Transformation of the Mobility System

\section{The Power of Mobile Sensing and Data Inferences to Support Societal Change}

Nowadays ubiquitous devices such as smartphones and -watches accompany their users throughout the whole day. We envision the following technologies as means to support societal change and implement approaches pointed out by environmental-psychology research in the previous section.

\paragraph{Ubiquitous Behavioral Data}
With mobile sensing methods, these devices can access data on the user's behavior, context, and situation unobtrusively in the background \cite{harari2020sensing,harari2017smartphone}. Common behavioral data encompasses but is not limited to device usage, and mobility behavior including the choice of means of transport (e.g. via Google's Awareness API \footnote{\url{https://developers.google.com/awareness/overview}, last accessed 7th February 2023}), and mobile language use. Information on behaviors that cannot be directly sensed by the smartphone, such as consumption and nutrition behaviors, can either be gathered with journaling methods \cite{van2017experience} (e.g. asking the user daily for their consumed amount of meat), via third-party devices or services (e.g. financial APIs that have access to purchases), or a semi-automatic approach combining both (e.g. taking a photo of each meal that is processed by image recognition) \cite{bemmann2020self}. Most data is available immediately in the situation (in situ), allowing the user to follow their progress live.

\paragraph{Machine Learning based Inferences}
Making inferences from behavioral data further allows assessing non-directly measurable behaviors and attitudes, such as personality traits \cite{stachl2020predicting} and political orientation \cite{khatua2020predicting}. Explained decisions of models support users in reflecting on their data and identifying connections between and reasons for behaviors \cite{bemmann_2022_piandai}.
%\todo{sth else that is closely related? exists research predicting environmental attitude?}
%\todo{One two more sentences why this is cool}

\paragraph{Mobile Crowd Data}
Data becomes especially powerful when it is put into context, i.e. comparing it with one's own historical data or with the data of others. Data of other groups of people can be collected either via mobile crowd sensing systems \cite{ganti2011mobile}, derived from existing sensing datasets of past studies such as conducted by \citet{schoedel2020basic}, or accessed via APIs. % such as \todo{national global behavior data APIs}.
Such comparisons can help people to classify their behavior with the local/national/global average. People can thereby also be pulled out of their bubble, which is a strong measure towards a sustainable attitude as depicted hands-on in \autoref{sec:extrap}. 

\section{Application Concepts}

In this section we interconnect the presented insights from environmental psychology with the specific capabilities of mobile sensing technology, to propose application concepts supporting societal change.

\subsection{Extrapolation of Sensed Behavior: Becoming Aware of Own Behavior}\label{sec:extrap}

\begin{figure}[t]
    \centering
    \includegraphics[width=\linewidth]{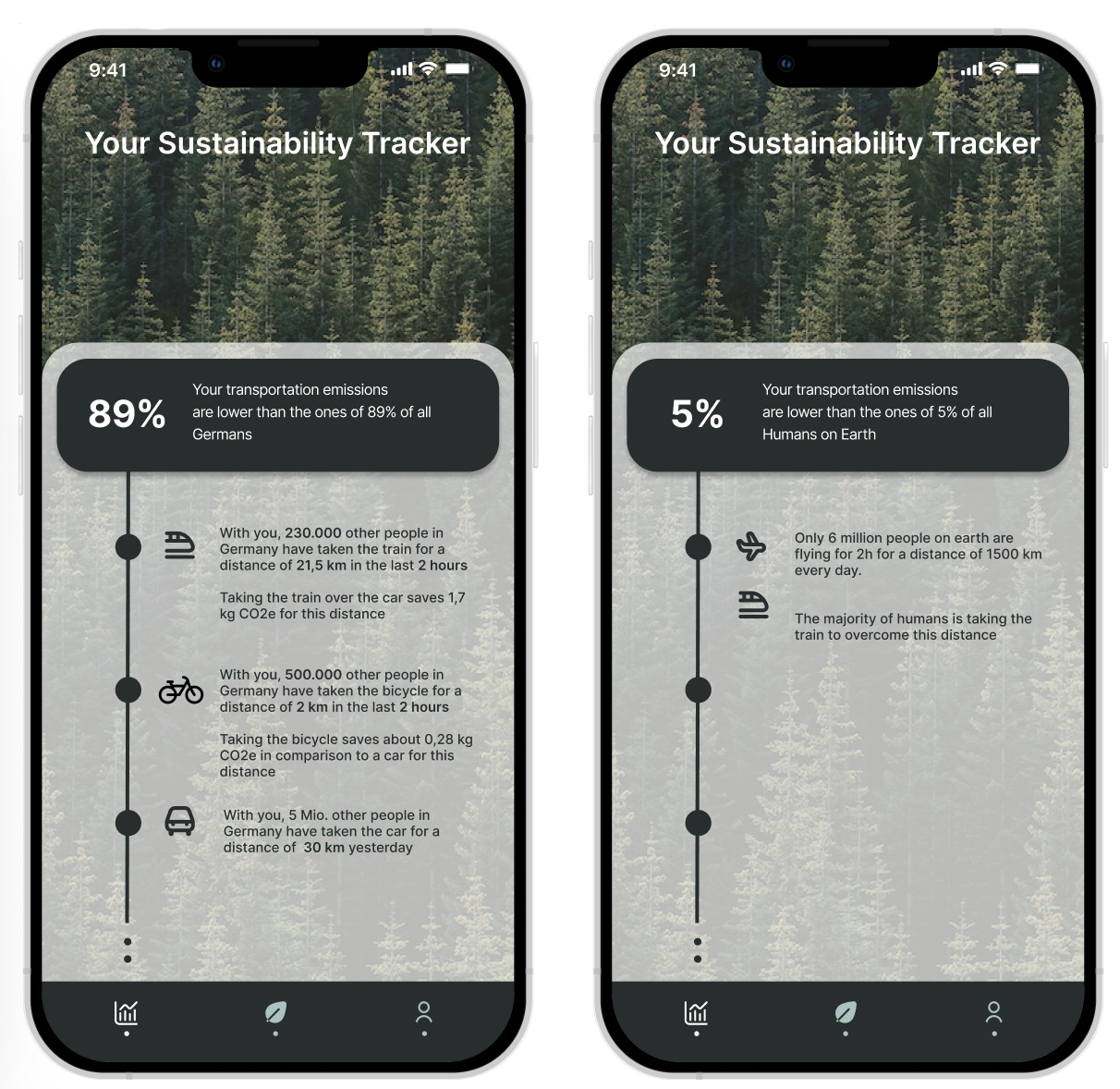}
    \caption{
    Our two proposed cases of extrapolated sensed behavior. In the application concept on the left side, comparrisons with a national average raise awareness for the common impact, see \autoref{sec:national-avg}. The right side helps users to arrange with the global population, making one aware of that taking a flight is a privilege of a minority of people, see \autoref{sec:global-avg}.}
	\label{fig:extrapolation-sketches}
\end{figure}

%\todo{= bewusst / Bewusstseinsbildend}

%#1 impact von handlung auf viele leute berehcnet aufzeigen
%- eigenes behavior sensen (z.b. mobilität und essen)
Many behaviors that have an ecological positive or negative impact can be captured with smartphone sensing in situ, i.e. at the moment when it happens. 
%\todo{Duplicate Mobile Sensing Intro: Entweder hier, oder oben bei den HCI powers, nicht beides so detailliert}
Data on environmentally-relevant behaviors, such as mobility or consumption, can be used by applications to track their progress over time, or support behavior change \cite{stawarz2015beyond}. 

A major factor limiting the proliferation of HCI towards sustainable behavior is the individual feeling of not having a higher-level impact.
%\todo{[REF to that shows that peope feel low impact of their efforts]}
This limits intrinsic motivation and post-use evaluations, leading to non-adoption of technology.
%, and technology paternalism \cite{huber2015gamification}. 

%
To overcome this issue, we envision an application that makes users conscious of their behavior in relation to others.
% Put behavior into the context of 
% ------ a) comparable peers (e.g. national, my community) -------
% zeigen wie viel Differenz hochgerechnet auf alle ausmacht: Wenn alle so leben wie du... In der Deutschland bubble wäre das eine Einsparung, in der worldwide Bubble wahrscheinlich eine Vershclechterung
\paragraph{Show environmental impact if everybody in your country behaves as you at the moment}\label{sec:national-avg} By taking the difference of the user's behavior to national average values
%\todo{put possible data sources in footnote, if available}
, users could be made aware of which impact one has as part of a larger group. By distinguishing between people that (a) already take efforts to live environmentally friendly and (b) those who don't, it could be further pointed out which impact it would have if (a) engaged individuals would stop their engagement (corresponding to lacking motivation) and (b) further people could be convinced. 
%result
This might foster a global identity feeling, which is a key factor to environmentally sustainable behavior \cite{reese2016common}.
% ----- b) global ------
% - man vergleicht sich nicht mehr mit seiner bubble ("aber alle die ich auf Insta sehe fliegen ja auhc viel") sondern mit der ganzen Welt, wo nur ein Bruchteil der Menschen fliegt
\paragraph{Show environmental impact if everybody in the world behaves as you do at the moment}\label{sec:global-avg} A different effect might be achieved when comparing with global averages 
%\todo{put possible data sources in footnote, if available}
. From the viewpoint of members of western societies, even the behavior of environmentally engaged people is carbon intensive when compared with the global average. The awareness of this should hint people to that (a) further engagement is still necessary, and (b) helps perceived losses of quality of life (e.g. renunciation of air travel) from outside their own bubble. While in one's (social media) bubble it seems usual to fly several times per year, this isn't the case when compared with the global standard. 
% result
This view should help users regard themselves as global citizens and to judge their behavior regarding global standards.

% general Umsetzung
% - app should be ambient / passive
\paragraph{General Design Considerations}
In general, such an application should be designed for passive use, i.e. the app giving the user information and food for thought occasionally when appropriate. Ambient narrative interfaces, such as visualization on the lock- and home screen as proposed by \citet{murnane2020designing}, are promising because users don't have to actively use them and research has shown that ambient information is easier to process \cite{ham2010ambient}. Also augmenting the real world, for example with public displays \cite{mathew2005using} or AR augmentations should be considered.

\subsection{Personality-based Targeting: Unconscious Attitude Formation}

%\todo{unterbewusst =>
%Vorteile:
%- rebound effkte (jetzt hab ich heute vegetarisch gegessen, dann kann ich morgne auto fahren) + climate depression
%- kann Leute ansprechbar die keine lust darauf haben}

\begin{figure}[t]
    \centering
    \includegraphics[width=\linewidth]{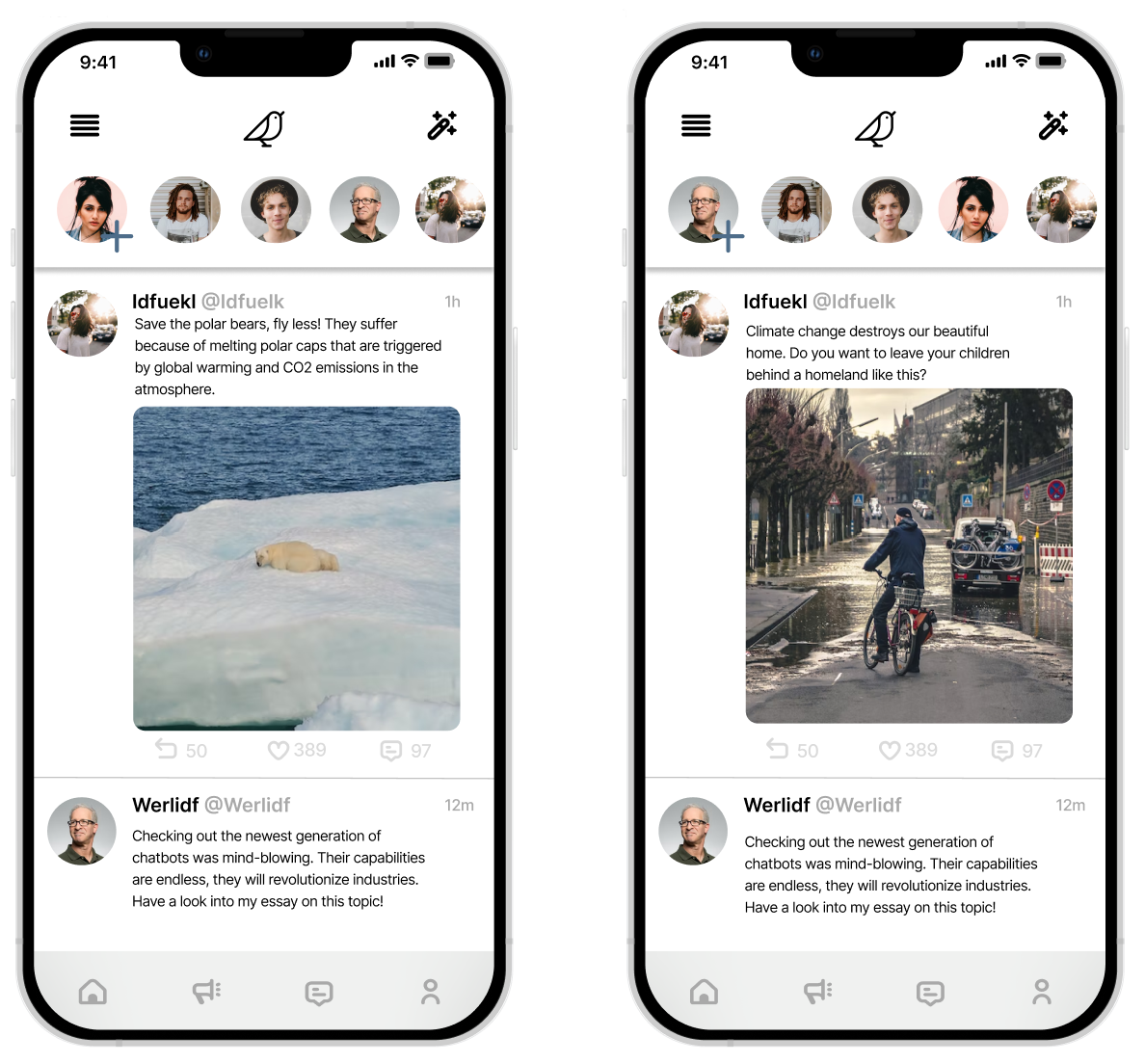}
    \caption{Psychographic messaging, targeting to two exemplary types of personality. Adapted from Cambridge Analytica's concept for targeting political campaigns with social media data\protect\footnotemark.}
    \label{fig:concerns-scenarios}
\end{figure}
\footnotetext{\url{https://mikefinnsfiction.wordpress.com/2017/09/26/how-cambridge-analytica-turned-social-media-into-political-weapon-for-hire/}, last accessed \today}

% #2 Leute zielgruppenabhängig ansprechen
% introduciton to psychometric targeting
Targeting content towards specific user groups has long existed especially in the context of the advertisement or election promotion campaigns, for example adapting ads by location \footnote{\url{https://www.ndi.org/sites/default/files/Module\%203_Research\%2C\%20Strategy\%20and\%20Targeting_EN.pdf}, last accessed 6th February 2023}, nowadays known as \textit{macro targeting}. With the rising availability of more detailed user data, targeting procedures became more personalized and dynamic. From targeting ads to situations (e.g. work vs. leisure \cite{banerjee2012location}) up to targeting content to an individual's personality, known as psychometric targeting \cite{matz2017psychological,dhawan2022outside}. These individual targeting mechanisms are also known as micro-targeting \cite{bennett2021understanding}. Micro-targeted ads unconsciously influence their audience, by speaking to fears and other subconscious triggers. Cambridge Analytica demonstrated the power of such technology, by influencing a.o. the Donald Trump election and Brexit vote \cite{cambridgeanalytica} with mass persuasion through targeted content based on social media data. 
% application for environmental chnage?
The border between clearly unethical use cases of psychometric targeting methods, such as the raisal of people's fear supported by fake news in the Donald Trump campaign, and societally accepted uses, such as personalized advertisements on social media recommending products in one's area of interest, is a continuum.
% continue: bis owhin kann man das für SHCI anwendne?
Research should discuss to which extent the application of psychometric targeting can also be used for the good in an ethical manner (continuing on e.g. \citet{bay2018social}).
% wie man auf Umweltverhalten targeten kann
%\todo{Welche Arten gibt es in RW Leute einzuteilen?}
%\todo{Was gibt es um Umweltattitude u.Ä. zu predicten?}
Barriers to sustainable behavior are diverse and depend on individual norms, education, and experiences. One's attitude can make an exemplary subdivision: Among people whose general attitude is in favor of sustainable behavior, the \textit{attitude-behavior gap} describes reasons that hinder actual sustainable behavior. On the other hand, there are people whose attitudes are not in favor of acting sustainably at all. 
Both groups of people have to be targeted differently when designing systems supporting sustainable behavior. In the first case, it is promising to support people in their intended actions (e.g. lowering burdens of the behavior). However, in the latter case, persuasion of one's internal beliefs and attitude would have to go first.
% based on which data?
%\todo{based on which data?}
% What to target
Targeting could encompass various kinds of content. Advertisements and pro-environmental campaigns in social media could be targeted, to approach the viewer's individual burden against sustainable behavior. The unconscious approach could thereby bypass limitations of conscious targetings, such as rebound effects and climate depression %\todo{ref for climate deprseison and maybe rebound effects? (hoewver the latter is common knowledge imo...)}. 
Furthermore, it enables us to talk to audiences that are not inherently interested in the topic of climate change.
%Webbrowser and search engines could augment environmental choices, as the search engine Ecosia does for some web pages \footnote{\url{https://blog.ecosia.org/green-search/}, last accessed 6th of February 2023}. 

%\todo{Es reicht wenn 1 oder 2 Prozent der Leute daten für ein Profiling sytsem bereitstellen, damit die profile funktioinerne}

%Vorteile:
%- rebound effkte (jetzt hab ich heute vegetarisch gegessen, dann kann ich morgne auto fahren) + climate depression
%- kann Leute ansprechbar die keine lust darauf haben

% \subsection{Real World Augmentation of Consequences + Handlungsmöglichkeiten}

% \todo{Diese Section ggf weglassen (nicht mein Kerntopic) und in Discussion als Ausblick zur Visualisierung von 3.1 nennen}

%\section{#3}
%Identitätsgefühl
%Wie kann man ein "globales Identitätsgefühl" mit Technologie fördern?
% -> auch in #1

%\section{#4}
%was mit social media machen? Leute über ihre bubble hinausgucken lassen? (angneommen man wäre facebook...)
% -> das mit der bubble ist auch in 1 mit drin

% #### Discussion ####
\section{Discussion}

%\todo{Keep this seciton short but thought provoking}

\subsection{Why should one use such a system?}

The usage of technology mostly happens deliberately, i.e. users decide to use it in expectation of some benefit. As extrinsic motivators are often not present in this domain, and intrinsic factors hardly overcome other aggravating factors, HCI needs to find solutions motivating the use of sustainability-fostering application concepts. We'd like to discuss the potential of ambient mobile applications and public displays in the workshop. Also, the involvement of third-party stakeholders who are interested in an environmentally friendly attitude should be considered when designing applications, such as governments and pro-environmental parties.
%\todo{Parteien sind daran ineressiert die Attidue von Leuten in ihre Richtung zu ändern (könnten Geld invetsiern)}

%(however as Ecosia actively has to be chosen, it is only used by people with general pro-environmental attitude)

\subsection{Ethical Considerations of (Mis)using Technology}

Technology brings a lot of power to their developers. Psychometric targeting approaches have played a major, if not deciding, role in the election of Donald Trump as U.S. president and the Brexit vote. We would like to discuss in the workshop whether the application of such technologies for the common good is ethically correct.

\subsection{Sustainable = Good?}

What is a \textit{good} purpose is a matter of perspective. For the audience of this paper, it might be undebatable that fostering sustainability is a good aim and political popularism isn't. However outside of this bubble, for example from the viewpoint of a confident republican politician, it might be vice versa.
As a basis for the previous discussion point, we need to discuss whether what is \textit{good} can be defined at all. Is it ethically correct to try to convince people with our pro-environmental viewpoint?

%\todo{Zoom out: Wir geben vor dass umweltfreundlichkeit ein gutes Ziel ist. Ist das nicht wieder nur eigene Meinung und paternalism? ISt es ehtisch überhaupt korrekt Leute für umweltfreundliches Vehralten zu überzeugen?}

\subsection{Combination with Further Technologies}

The proposed concepts could be well-combined with other technologies. For example, Virtual Reality (VR) might be suitable to depict the effect of one's behavior in a future world, or Augmented Reality (AR) could augment alternative behaviors in situ.

%In der "brennenden AR Welt" dann Hanldungsoptionen einblenden. Das wäre dann ein Bildungstool - TODO in Struktur einarbeiten

%passt zu "making data understandable, usable and actionable"
%\todo{Idee von Carmen: Zum ersten Bereich „making data understandable, usable and actionable“ ist mir direkt Prof. Pongratz (link) eingefallen da sie sich ziemlich gut mit Earth system models auskennt, gerade solchen die im IPCC Report verwendet wurden. Da besteht echt noch einiges an Potenzial das noch greifbarer darzustellen. Sie wäre bestimmt dabei darüber zu sprechen wie man die Daten aus den Modellen aus einer HCI Perspektive besser darstellet.}

%VR-Zeitmaschine?

%mountain game: http://mountain-game.com/

\section{Conclusion: Next Steps}

Beyond reducing its own impact and making technology more efficient, HCI is having a limited impact on solving the climate crisis. Societal and individual change is essential, as the people are those who steer businesses and governments with their consumption respectively election decisions.

To come up with novel approaches, we argue for more interdisciplinary work in this domain. We think that combined insights and joined thoughts of HCI, (environmental-)psychology, sociology, and other fields are promising. As the next step, we envision conducting an interdisciplinary workshop.
%- Interdisciplinary workshops! Knoweldge has "just" be put together; was successful in our pssychology context

\bibliographystyle{ACM-Reference-Format}
\bibliography{sample-base}


\end{document}